\documentclass[preprint]{aastex}
\input epsf

\newcommand{\msun}{\mbox{M$_{\odot}$}}
\newcommand{\ishape}{{\tt ishape}}

\newcommand{\viz}{\mbox{$(V\!-\!I)_0$}}
\begin{document}
\title{A G1--like globular cluster in NGC~1023
 \footnote{Based on observations obtained with the NASA/ESA Hubble Space 
 Telescope, obtained at the Space Telescope Science Institute, which is 
 operated by the Association of Universities for Research in Astronomy, 
 Inc. under NASA contract No.\ NAS5-26555.}
}

\author{S{\o}ren S. Larsen
  \affil{UC Observatories / Lick Observatory, University of California,
         Santa Cruz, CA 95064, USA}
  \email{soeren@ucolick.org}
}

\begin{abstract}
  The structure of a very bright ($M_V=-10.9$) globular cluster in NGC~1023 
is analyzed on two sets of images taken with the Hubble Space Telescope. 
From careful modeling of King profile fits to the cluster image, a core radius 
of $r_c = 0.55\pm0.1$ pc, effective radius $R_e = 3.7\pm0.3$ pc and a central 
surface brightness of $\mu_0(V) = 12.9\pm0.5$ mag arcsec$^{-2}$ are derived.
This makes the cluster much more compact than $\omega$ Cen, but very 
similar to the brightest globular cluster in M31, G1 = Mayall II.  The 
cluster in NGC~1023 appears to be very highly flattened with an ellipticity 
of $\epsilon \approx 0.37$, even higher than for $\omega$ Cen and G1, and
similar to the most flattened clusters in the Large Magellanic Cloud.
\end{abstract}


\section{Introduction}

  A very bright globular cluster in the nearby lenticular galaxy NGC~1023 was 
recently identified by \citet{lbk01}.  At a distance of $9.9\pm0.6$ Mpc 
\citep{cia91}, NGC~1023 is close enough that images taken with the 
\emph{Hubble Space Telescope} (HST) will show globular clusters (GCs) as 
spatially extended objects, and the structure of individual clusters with 
sufficient S/N can be studied in considerable detail. It is of interest to 
compare the structural parameters (size, ellipticity etc.) of globular 
clusters in different galaxies because significant differences are known to 
exist even among GCs in the Local Group. For example, clusters in the Large 
Magellanic Cloud (LMC) are more flattened than those in the Milky Way 
\citep{gh80}. It has also been noted that the \emph{brightest} GCs in the 
Milky Way and M31 ($\omega$ Cen and G1 = Mayall II, respectively), as well
as in the Magellanic Clouds, are the most flattened in their respective
host galaxies
\citep{van84}, and \citet{van96} has suggested using the HST to test if
the brightest globular clusters around other galaxies are also flattened.
In the present paper the structure of the bright globular cluster in 
NGC~1023 is analyzed using archive HST images, and it is then compared 
with $\omega$ Cen and G1. 

%
%
%
%
%
%
%
%

\section{The bright cluster n1023-13}

  \citet{lbk01} presented spectroscopy of 11 old globular clusters in 
NGC~1023, most of which were selected from WFPC2 exposures obtained as 
part of Program 6554 (PI: Brodie) and discussed in \citet{lb00}. However,
a few additional objects were selected from ground-based images, one 
of which (n1023-13) turned out to be a very bright globular cluster.
The 2000.0 coordinates are $\alpha$ = 2:40:27.84 and $\delta$ = 39:04:40.2.
Although this object is close to the saturation limit on the Brodie dataset
and located near the edge of one of the WF chips, it is also included on 
a shorter WFPC2 exposure of NGC~1023, obtained for D.\ Richstone (HST 
Program ID 6587).  This dataset consists of 5 integrations of 260 sec 
each in the F555W band and 2 integrations of 900 sec in the F814W band, 
short enough that the cluster is well below the saturation limit.

  Fig.~\ref{fig:images} shows a close-up of the Richstone WFPC2 image of 
the cluster, compared to a star of similar brightness. The cluster is clearly 
resolved and appears to be significantly elongated.  At a distance of 9.9 Mpc,
one WFPC2 pixels spans 4.8 pc, comparable to or somewhat larger than the 
typical half-light radius of a globular cluster \citep[e.g.][]{kw01},
so special care must be taken when analyzing the light profile of clusters
at this distance. Nevertheless, the S/N of n1023-13 is sufficient 
to allow a relatively detailed analysis of its structure and compare with 
other bright globular clusters like $\omega$ Centauri in the Milky Way and 
G1 = Mayall II in M31.  

\subsection{Modeling of the light profile}

  Of the 5 available F555W exposures in the Richstone dataset, only two 
were used for the analysis presented here. The two selected exposures
required no shifts in the $x$ and $y$ directions, and consequently no 
resampling of the Point Spread Function (PSF), before combination. 
Addition of the individual images, including elimination of cosmic ray
hits, was done with the IMCOMBINE task in IRAF\footnote{IRAF is distributed 
by the National Optical Astronomical Observatories, which are operated by 
the Association of Universities for Research in Astronomy, Inc.~under 
contract with the National Science Foundation}.
Inspection of the data quality files for the Brodie WFPC2 image showed 
that the image of n1023-13 was, in fact, not saturated, so we are in the 
fortunate situation of having two WFPC2 images of the same cluster, 
allowing for a useful consistency check of the results.  The cluster is 
located on two different WF chips (WF2 on the Richstone and WF4 on the 
Brodie dataset), and at very different positions within the chips 
($(x,y) = 210, 560$ versus $(x,y) = 740, 79$) in the two pointings.
  
  The F555W images of the cluster were modeled using the \ishape\ algorithm 
\citep{lar99}, assuming various analytical models for the intrinsic 
luminosity profile of the cluster and then convolving with the HST 
PSF, generated by the TinyTim software \citep{kri97}. 
A convolution with the WFPC2 ``diffusion kernel'' was also included. 
The ellipticity, Full Width At Half Maximum (FWHM) and orientation of the 
model profiles were iteratively adjusted until the best match with the 
observed image was obtained.  As the TinyTim manual and \emph{WFPC2 handbook}
give two slightly different diffusion kernels, fits were done for both of them. 
Because the diffusion kernel is wavelength dependent but only given for the 
F555W filter in the manuals, no fits to the F814W images were attempted.

 Two different types of analytical profiles were assumed: 
\citet{king62} profiles of the form
\begin{equation}
  \mu(r) = k\left[\frac{1}{\sqrt{1 + r^2/r_c^2}} 
                - \frac{1}{\sqrt{1 + r_t^2/r_c^2}}\right]^2, \; r<r_t
  \label{eq:king}
\end{equation}
and ``Moffat'' profiles given as
\begin{equation}
  \mu(r) = \mu_0 \left[1+(r/r_c)^2\right]^{-\alpha}.
  \label{eq:moffat}
\end{equation}
The King profiles are characterized by a core radius $r_c$ and a tidal
radius $r_t$ and are known to provide excellent fits to globular cluster 
luminosity profiles. The Moffat profiles are similar to the profiles used 
by \citet{els87} to fit young LMC clusters, but note that the $\alpha$ 
in Eq.~(\ref{eq:moffat}) is equivalent to their $\gamma/2$. In the limiting 
case of $\alpha=1$, Eq.~(\ref{eq:moffat}) is identical to a King profile with 
infinite tidal radius.

  The version of \ishape\ used here allows the concentration parameter 
$c=r_t/r_c$ of the King 
profiles and the exponent $\alpha$ of the Moffat profiles to be kept at a 
fixed value, or to vary these parameters during the fitting process.  The
$c$ parameter is the most uncertain of the fitted parameters, and the
exact value returned by \ishape\ was quite dependent on the initial
guesses.  Generally, the fitted $c$ values were between 100 and 300.
In order to ensure better stability in the fits and assess the effect of
variations in $c$, the remaining shape parameters (ellipticity, FWHM, 
orientation) were thus determined for fixed $c$ values of 100, 200 and 300.
The Moffat fits, on the other hand, were quite stable and consistently
returned $\alpha$ values around $1.20\pm0.05$.


  The fitting radius was set to 15 pixels or $1\farcs5$, corresponding to a 
radius of 72 pc for a distance of 9.9 Mpc. A number of fits were also done
for a fitting radius of 25 pixels, but for this larger radius the background
gradient becomes noticeable, although the results of the fits remained 
essentially identical to those done with the smaller radius. 

  Figure~\ref{fig:fitfig} shows the results of various fits to the
Richstone image. From the left: King 
profiles with $c=10, 30, 100, 300$ and $\infty$ and the best-fitting Moffat 
profile with $\alpha=1.20$. The top row shows the best-fitting model images
generated by convolution of the King/Moffat profiles with the TinyTim 
PSF and diffusion kernel, and the bottom row shows the residuals when
the model images are subtracted from the observed image. The King profiles
with $c<100$ or $c=\infty$ are clearly unable to reproduce the observed 
cluster image, while the $c=100$ and $c=300$ profiles show little
structure in the residuals beyond the photon noise. Formally, the
best fit (lowest $\chi^2$) was obtained for $c\sim200$.  Increased S/N 
could potentially help in constraining the King profile $c$ parameter and
for this purpose a number of fits were carried out on all 5 Richstone images 
combined.  These tended to converge towards $c\approx150$. It thus seems 
reasonable to adopt $c=200\pm100$ for the concentration parameter. The 
$\alpha=1.2$ Moffat profile also provides a very good fit, but the King 
profile fits which are physically better motivated will be used for most 
of the discussion in this paper.

  The various King and Moffat profiles are plotted in Fig.~\ref{fig:profs},
scaled to the same luminosity within 1\farcs5 and computed for the actual
fitted core radii. Note that Fig.~\ref{fig:profs} shows the \emph{intrinsic}
cluster profiles, not the observed ones (which are obtained by convolving
the profiles in Fig.~\ref{fig:profs} with the HST PSF and diffusion kernel).
Since 1 WF pixel spans 0\farcs1, it is clear that the
central parts of the profiles are unresolved and the differences in the
fits rely mainly on the behaviour in the wings. 

\subsection{Shape parameters for best fitting King profiles}

  The best fit parameters are given in Table~\ref{tab:fitpar} for each
of the two WFPC2 datasets. Each value in the Table is an average of
6 individual fits, for King profiles with $c=100, 200$ and 300, and
for the two different diffusion kernels. The quoted errors are simply
the standard deviation of these individual measurements, which is probably
a better estimate of the true uncertainty than the standard error on
the mean.

  The first rows in the Table give the photometry in a 0\farcs5 (5 pixels)
aperture, corrected for reddening towards NGC~1023.  The photometry was done
with the PHOT task in DAOPHOT directly on the images, measuring the background 
as the mode of all pixels in an annulus starting at 20 pixels and 10 pixels 
wide.  Calibration to standard $V$ and $I$ magnitudes was done following 
\citet{hol95} and using the \citet{sch98} extinction value of $A_B = 0.262$ 
mag. The distance modulus of $29.97\pm0.14$ derived by \citet{cia91} was based 
on \citet{bh84} extinctions, which often tend to be somewhat lower than the 
\citet{sch98} values. However, for NGC~1023 the two happen to agree to within 
0.01 mag in $A_B$.  

  For an extended object like n1023-13, the $0\farcs5$ aperture slightly 
underestimates the total brightness. According to \citet{hol95}, the WFPC2 
PSF itself scatters about 10\% of the light to radii beyond their $0\farcs5$ 
reference aperture, while measurements of n1023-13 through an $r=20$ pixels 
aperture gave total $V$ magnitudes brighter by about 0.2 mag compared to 
the $r=5$ pixels measurements. Thus, the $V$ magnitudes in 
Table~\ref{tab:fitpar} are probably too faint by $\sim 0.1$ mag, and after
applying this correction the absolute magnitude becomes $M_V=-10.9$.
Likewise, direct integration of the King profiles also yields a
correction of about 0.1 mag from $r=0\farcs5$ to $r=\infty$.

  There is excellent agreement between the photometry from the two pointings,
as well as between the \ishape\ profile fits.  The two sets of fits give 
FWHM values of $0\farcs027\pm0\farcs006$ and $0\farcs029\pm0\farcs007$, 
minor/major axis ratios of $0.62\pm0.02$ and $0.64\pm0.01$, and the position 
angles on the sky agree to within 3 degrees.  Note that the relatively 
large range in the $c$ parameter does not lead to correspondingly large 
uncertainties on the fitted FWHM.

  The FWHM and concentration parameter returned by \ishape\ can 
be converted to more familiar quantities such as the core radius ($r_c$)
and half-light (effective) radius ($R_e$).  For a King profile the FWHM 
and $r_c$ are related as
\begin{equation}
  \mbox{FWHM} = 2 \left[\left(\sqrt{1/2} 
                            + \frac{1-\sqrt{1/2}}{\sqrt{1+c^2}}\right)^{-2}
                         -1\right]^{1/2} \, r_c,
\end{equation}
i.e.\ FWHM $\approx 2 \, r_c$ for $c\gg1$.
A similar simple analytical relation between the FWHM or $r_c$ and $R_e$ does 
not exist, but can be approximated by a power-law of the form
\begin{equation}
  R_e / r_c \approx 0.547 \, c^{0.486} 
  \label{eq:rerc}
\end{equation}
For $c>4$ this approximation is good to $\pm2$\%.

These relations assume that the profiles are circularly symmetric, while
the FWHM values returned by \ishape\ are measured along the major axis
of the fitted profiles.  In order to compute the core and effective radii in 
Table~\ref{tab:fitpar}, an average of the major and minor axis FWHM has
been used. This results in core radii of $r_c = 0.52\pm0.12$ pc and 
$r_c = 0.57\pm0.13$ pc for the two datasets, and effective radii of 
$R_e = 3.47\pm0.33$ pc and $R_e = 3.85\pm0.34$ pc, respectively, at the 
assumed distance of NGC~1023.

\subsection{Ellipticity}

  The minor/major axis ratio returned by the fits is quite stable
at around $0.63\pm0.01$, corresponding to a very large ellipticity of 
$\epsilon = 0.37\pm0.01$.
An independent estimate can be obtained by using the ELLIPSE task in
the STSDAS package to fit elliptical isophotes directly to the cluster
image, although these will tend to underestimate the ellipticity 
because the true shape is blurred by the HST PSF. This effect will be
worse closer to the center, while ellipse fits at large radii are
uncertain because of low S/N. At $r=5$ pixels, the ELLIPSE task
returns ellipticities of $\epsilon = 0.24$ on both the Richstone F555W 
and F814W exposures, and 0.14 and 0.20 on the Brodie F555W 
and F814W exposures.  Irregularities in the background due to the proximity 
to the edge of the image are clearly visible in the Brodie exposures and
the ELLIPSE fits are probably less accurate for these data.

  ELLIPSE fits were also done on the model images generated by \ishape\
(Figure~\ref{fig:fitfig}).  These fits returned an ellipticity of 
$\epsilon=0.25$ at $r=5$ pixels, in nearly perfect agreement with the fits 
to the actual science images. Thus, it is clear that the blurring of the 
cluster image by the HST PSF leads to an underestimation of $\epsilon$ 
when measured directly on the images, and the true answer probably lies 
closer to the $\epsilon\sim0.37$ value from the King profile fits.

\subsection{Central surface brightness}

  Although the central parts of the King profiles are unresolved on the
WFPC2 images, the central surface brightness for the 
cluster can still be computed, assuming that the King profile continues all 
the way to the center. The central surface brightness $\mu_0$ of a 
\citet{king62} profile and the total luminosity $L(R)$ within a radius 
$R < r_t$ are related as
\begin{equation}
  \mu_0 = k \left[1-\frac{1}{\sqrt{1+c^2}}\right]^2
\end{equation}
and
\begin{equation}
  L(R) = \pi k \left[r_c^2 \ln \left(1+\frac{R^2}{r_c^2}\right)
                     + \frac{R^2}{1+c^2}
		     - \frac{4 r_c^2}{\sqrt{1+c^2}}
		       \left(\sqrt{1 + \frac{R^2}{r_c^2}} -1 \right)
                \right]
\end{equation}
For reddening-corrected $V=19.19$ and $V=19.17$ within a radius of 
$R=0\farcs5$, this results in $\mu_0 = 12.8\pm0.4$ mag arcsec$^{-2}$ and 
$\mu_0 = 13.0\pm0.4$ mag arcsec$^{-2}$ for the two fits. Again, the
errors are computed as the standard deviation of the 6 values based on the 
individual fits.  The computed $\mu_0$ values are not very 
sensitive to the exact concentration parameter and most of the uncertainty 
comes from the core radius. Alternatively, one can also compute
the central surface brightness for the Moffat profile fit. This yields
$\mu_0 = 13.6$ mag arcsec$^{-2}$, somewhat fainter than for the King profile.  
However, a King profile is probably a better approximation to the true 
light profile of the cluster and a value around $\mu_0 = 12.9\pm0.5$ mag 
arcsec$^{-2}$ may be a reasonable estimate of the central surface 
brightness and its associated error.

\section{Comparison with G1 and $\omega$ Cen}

  Table~\ref{tab:fitpar} also lists a number of properties for two other 
bright globular clusters: $\omega$ Cen, the brightest GC in the Milky 
Way, and G1 = Mayall II, the brightest known GC in M31 \citep{me53}. 
\citet{pv84} first noted a significant 
flattening of G1 and estimated an ellipticity of $\epsilon=0.22$.  Using HST 
images of G1, \citet{rich96} measured a core radius of 
$r_c = 0\farcs170\pm0\farcs011$ or $0.63\pm0.04$ pc for a distance modulus
of 24.42 \citep{fm90} and found a tidal radius of $28\farcs2$ = 
105 pc, corresponding to a concentration parameter $c=166$ and an effective 
radius of 4.1 pc.  They obtained a central surface brightness of 13.5 mag 
arcsec$^{-2}$ in $V$ and a slightly higher ellipticity 
($\epsilon=0.25\pm0.02$) than the one reported by \citet{pv84}.  Also using HST
images, \citet{mey01} measured a somewhat smaller core radius of 
$r_c=0.52$ pc, but a larger tidal radius of 200 pc for G1. They quote a 
\emph{half-mass} radius of $r_h=14$ pc, corresponding to an effective radius 
of 10 pc \citep[p. 16]{spi87} if light traces mass, but their 
Table 3 indicates an effective radius closer to $1\farcs18$ or 4.4 pc, in 
better agreement with Equation~(\ref{eq:rerc}) which gives
$R_e = 0.52 \, {\rm pc} \times 0.547 \, (200/0.52)^{0.486}$ = 5.1 pc. Contrary
to \citet{rich96}, \citet{mey01} found a variation in ellipticity with 
radius, ranging from $\epsilon=0.1$ in the innermost and outermost parts, to 
$\epsilon=0.3$ at a radius of $2\farcs1$ or 8 pc, with a mean of 
$\epsilon\sim0.2$. All of the above studies agree on an integrated $V$ 
magnitude of $M_V=-10.9$.

  Most of the relevant parameters for $\omega$ Cen are listed in the McMaster 
catalogue \citep{har96}.  This cluster has $M_V=-10.24$ and is also quite 
elongated with $\epsilon = 0.19$ \citep{ff82}. It is, however, more extended 
than G1 and has a core radius of 3.8 pc.  The effective radius is 6.2 pc, 
somewhat larger than the typical $3-4$ pc for globular clusters.

  Compared to these two bright globular clusters in the Local Group,
n1023-13 appears to be nearly an identical twin of G1.  Both clusters are 
of comparable luminosity and their effective radii are quite similar to 
those for globular clusters of much lower luminosities. This is consistent
with the lack of correlation between effective radius and cluster
luminosity observed in the Milky Way \citep{dm94}.  Both G1 
and n1023-13 have very compact cores and, as a result, very high central 
surface brightnesses.  In fact, the highest central surface brightness for any 
Galactic globular cluster is $\mu_0 = 14.15$ (NGC 1851, from the McMaster 
catalog) and $\omega$ Cen itself has a much lower 
$\mu_0 = 16.8$ mag arcsec$^{-2}$.

  From the core radius and central surface brightness, the central density 
$\rho_0$ can be estimated from the relation
\begin{equation}
  \rho_0 =
  \frac{3.44\times10^{10}}{P r_c}
   10^{-0.4 \mu_0(V)} (M/L) \, \msun \, \mbox{pc}^{-3}
   \label{eq:dens}
\end{equation}
with $P\approx2$ and $r_c$ in pc \citep{pk75,wb79}.  For G1 and n1023-13 
this leads to $\rho_0 = 2.0\times10^5$ M$_{\odot}$ pc$^{-3}$ and 
$\rho_0 = (4.1\pm2.4)\times10^5$ M$_{\odot}$ pc$^{-3}$, respectively, assuming 
a M/L ratio of 1.6 \citep{il76}. Here we have used the average of the two 
estimates of $\rho_0$ for n1023-13 listed in Table~\ref{tab:fitpar}. From 
King-Michie models, \citet{mey01} obtain a central density of $\rho_0 = 
4.7\times10^5$ M$_{\odot}$ pc$^{-3}$ for G1, in fair agreement with the
estimate obtained from Eq.~(\ref{eq:dens}). Again, n1023-13 has a similar 
high central density to G1, while $\omega$ Cen has a central density of only 
$\sim1400$ M$_{\odot}$ pc$^{-3}$.

  The metallicity of $\omega$ Cen is given as [Fe/H] = $-1.62$ in the McMaster
catalogue. Metallicity estimates for G1 range between $-1.19\pm0.25$
\citep{bh90} and $-0.7$ \citep{rich96}, while \citet{mey01} find an
intermediate value of [Fe/H] = $-0.95\pm0.09$.  For n1023-13, \citet{lbk01} 
found [Fe/H] = $-1.15\pm0.22$ from Keck spectroscopy, and a very similar 
value of [Fe/H] = $-1.11$ from the $V-I$ color. n1023-13 thus seems to be 
somewhat more metal-rich than $\omega$ Cen, but of comparable metallicity 
to (or maybe slightly lower than) G1. 
It is also worth noting that -- unlike most globular clusters -- both 
$\omega$ Cen and G1 have quite a large internal scatter in their metallicity 
distributions. Whether or not this is the case for n1023-13 is, however,
impossible to tell from the current data.

  One common property of all these bright globular clusters is their
high ellipticities. The \ishape\ fits indicate an ellipticity as high
as $\epsilon \approx 0.37$ for n1023-13.  $\omega$ Cen has an ellipticity 
of $\epsilon = 0.19$, while G1 appears to display a range of ellipticities 
as a function of radius, reaching a maximum of $\epsilon = 0.30$ at 
intermediate radii. The present data do not permit an analysis of internal 
variations in the shape parameters (ellipticity, orientation) of n1023-13, 
but from the mean values it seems that the cluster may be even more 
flattened than $\omega$ Cen and G1. Is an ellipticity $\epsilon = 0.37$ 
unrealistically large? Although most globular clusters in the Milky Way have 
low ellipticities, clusters in the Large Magellanic Cloud tend to be
much more elongated. One of the more striking examples is the
intermediate-age ($\sim 2$ Gyr) cluster NGC~1978 with
an ellipticity of $\epsilon = 0.30$ \citep{fis92}. After correction for
projection effects, \citet{gh80} find that very few LMC clusters are
likely to be spherical, most having true ellipticities in the
range 0.2 -- 0.4. Thus, the high ellipticity for n1023-13 might not be 
unreasonable, especially if the cluster is seen nearly edge-on, but it
would definitely rank it as one of the most elliptical clusters known.

  Which mechanism might be responsible for the high ellipticities of 
massive globular clusters?  In a recent paper, \citet{cen01} has argued 
that many globular clusters may initially possess significant amounts of 
angular momentum, which should cause flattening of their initial shapes 
\citep{king61}.  Some evidence has been presented that rotation might 
indeed be a dominant 
mechanism in causing flattening of globular clusters \citep{dav86,ws87} and 
of $\omega$ Cen in particular \citep{mm86,mer97}.  On the other hand, 
\citet{fis92} found no evidence for rotation in the highly flattened 
cluster NGC~1978, although they could not rule it out.  However, it is 
not quite clear why this effect should affect the most massive clusters more 
strongly. Perhaps the tendency for more massive clusters to have longer 
relaxation times \citep[p. 40]{spi87} can cause an initial flattening to 
persist for a longer period of time in these systems, or maybe the more 
massive clusters tend to be born with larger angular momenta.  Detailed 
dynamical calculations and a better understanding of globular cluster 
formation will be required before these questions can be definitively 
answered.  

  Another mechanism which could produce elongated clusters with high angular 
momenta is merging of binary clusters \citep{sm89}. If the two clusters have 
different metallicities then the result might be an elongated cluster with 
a metallicity spread, as observed in $\omega$ Cen and G1, although it seems 
unlikely that two clusters born near each other (in time and space) would 
exhibit such large metallicity differences. Alternatively, the merger might 
result from a later chance encounter between two clusters \citep{ia88},
but this clearly makes the mechanism less attractive as a general explanation.

  Finally, \citet{mey01} have discussed the possibility that G1 and 
$\omega$ Cen are the remnants of nucleated dwarf galaxies. This might make 
the metallicity spread and elongated morphology of these two clusters more 
understandable, but does not appear to be a likely explanation for the 
high ellipticities of most LMC clusters.

\subsection{Sources of errors}

  Comparison of the two independent fits gives a good estimate of the random
errors involved in the fitting process. However, because the characteristic
sizes involved are on the same order of magnitude as the resolution of the
images, the fitted parameters are strongly dependent on correct modeling
of the WFPC2 PSF. One critical component is the ``diffusion kernel'', which 
takes into account charge diffusion between neighboring pixels in the CCDs. 
The diffusion kernel given in the TinyTim manual 
scatters less light to the surrounding pixels than the one in the 
\emph{WFPC2 handbook}, and this difference also influences the fits.
By comparison of the fits with the two different diffusion kernels, it
was found that the kernel given in the WFPC2 handbook generally gave
smaller FWHM values by about 0.02 -- 0.04 pixels or 
$0\farcs002 - 0\farcs004$. This is consistent with this kernel scattering 
more light into surrounding pixels, thus causing \ishape\ to fit a slightly 
narrower intrinsic profile.

  We also fitted the star shown in Fig.~\ref{fig:images} with and without 
the diffusion kernel.  When using the diffusion kernel, \ishape\ was unable 
to improve the King model fits relative to comparison fits obtained by using 
a delta 
function, as expected for an unresolved stellar image.  In contrast, if 
the diffusion kernel was omitted then \ishape\ clearly recognized the star 
as an ``extended'' object with an effective radius of about 0.2 pixels,
confirming that the ``raw'' TinyTim PSF underrepresents the size of
a point source and that the diffusion kernel needs to be taken into account.

  Errors in the distance to NGC~1023 are not included in the uncertainty
estimates in Table~\ref{tab:fitpar}. The $9.9\pm0.6$ Mpc value from
\citet{cia91} is based on the planetary nebula luminosity function, and
agrees well with the 10.2 Mpc distance estimate given by \citet{fab97}.
Size parameters (FWHM, $r_c$, $R_e$ and $r_t$) scale linearly with the 
distance, while the central surface brightness $\mu_0$ is 
distance-independent. Thus, from Eq.~\ref{eq:dens}, the central density 
is inversely proportional to the distance. Compared to the measurement 
errors, the uncertainty on the distance to NGC~1023 has only a minor 
effect on the results, even if the $\pm6$\% error estimate by \citet{cia91}
might be somewhat optimistic.

\section{Summary}

  Based on a detailed analysis of its spatial structure, the bright
globular cluster n1023-13 is found to be very similar to the G1 cluster in 
M31. Both have an absolute $V$ band magnitude of $M_V=-10.9$. Using the average 
values from Table~\ref{tab:fitpar}, the spatial profile of n1023-13 is well 
fitted by a King model with core radius $0.55\pm0.1$ pc, effective radius 
$3.7\pm0.3$ pc and concentration parameter $c\sim200$.  The cluster appears 
to be very flattened with an ellipticity around $\epsilon \approx 0.37$, 
making it one of the flattest star clusters known, and providing another 
example of the \citet{van84} observation that the brightest GCs in galaxies 
also tend to be the most flattened.  Assuming that the King profile 
continues all the way to the cluster center, a very high central surface 
brightness of $\mu_0(V) =12.9\pm0.5$ mag arcsec$^{-2}$ is derived, implying 
a central density of $(4.1\pm2.4)\times10^5$ \msun\ pc$^{-3}$. This is 
comparable to G1, but two orders of magnitude higher than for $\omega$ 
Centauri.

\acknowledgments

This work was supported by National Science Foundation grant number AST9900732.
I thank Jean Brodie and Sidney van den Bergh for comments on the paper 
and enlightening discussions, and the referee for useful suggestions.

\newpage
\onecolumn

\epsfxsize=140mm
\epsfbox{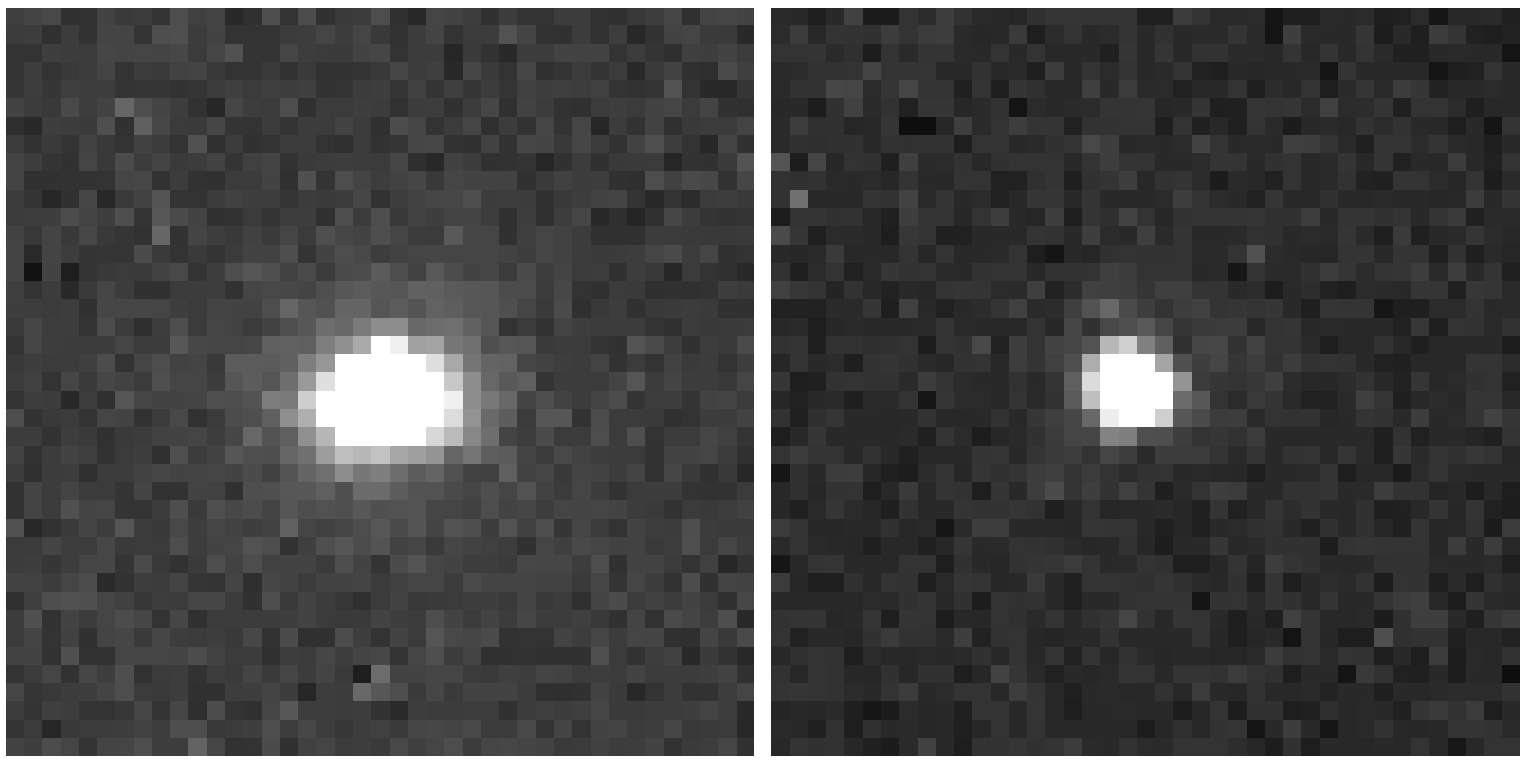}
\figcaption[Larsen.fig1.eps]{\label{fig:images}Images of 
the globular
cluster n1023-13 (left) and a star (right). The two objects are of about 
the same integrated magnitude and are shown with the same contrast settings. 
The cluster image is clearly more extended and appears to be elongated.
Each image spans $40\times40$ pixels or $4\arcsec\times4\arcsec$.
}

\epsfxsize=140mm
\epsfbox{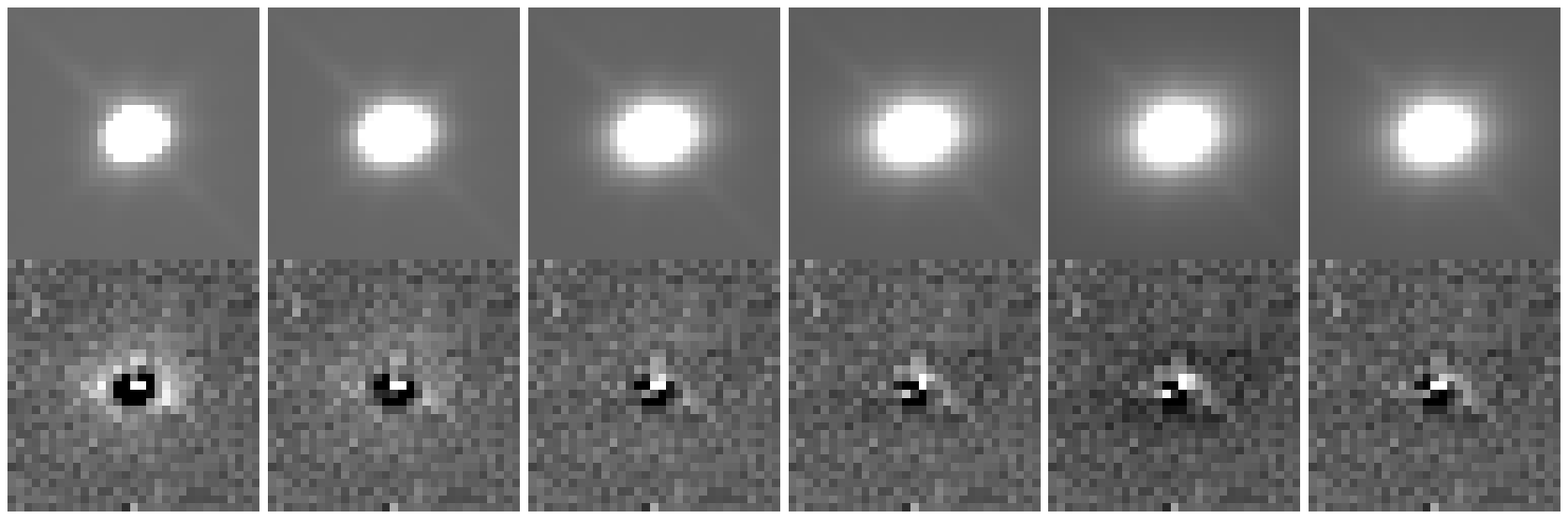}
\figcaption[Larsen.fig2.ps]{\label{fig:fitfig}Synthetic cluster images
and residuals for various analytical cluster models, fitted by
\ishape . From left to right: King ($c=10, 30, 100, 300, \infty$) and
Moffat ($\alpha=1.20$).
}

\epsfxsize=140mm
\epsfbox{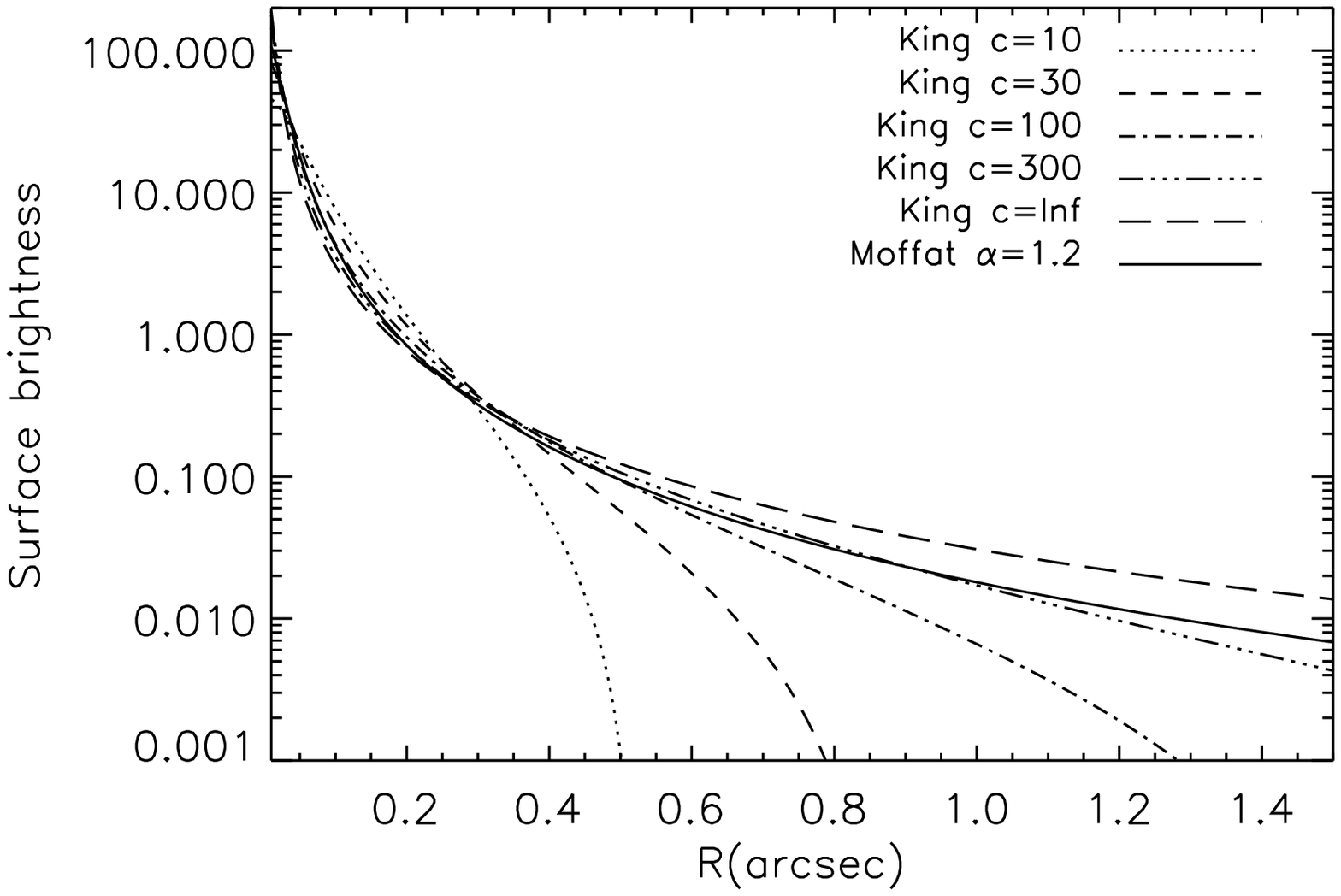}
\figcaption[Larsen.fig3.ps]{\label{fig:profs}
 Illustration of the various luminosity profiles fitted by \ishape .
 King models with concentration parameters $c=$ 10, 30, 100, 300 and $\infty$
 (dotted/dashed lines) and a MOFFAT profile with index $\alpha=1.20$
 (solid line).  All profiles have been normalized to the same luminosity 
 within $r=1\farcs5$. Note that these are the \emph{intrinsic} profiles,
 not taking into account the fact that the actual observed profiles are
 blurred by the WFPC2 PSF.
}
\newpage




\begin{deluxetable}{lrrrr}
\tablecaption{\label{tab:fitpar} Comparison of n1023-13, G1 and $\omega$
Cen.}
\tablecomments{
For $r_c$ and $R_e$ the given values are averages of minor and major
axis values. $V_0$ and $\viz$ are given for a $0\farcs5$ (5 pixels)
aperture. \\
$c$ = King profile concentration parameter = $r_t / r_c$.
P.A. = Position angle on sky, measured N through E.
$\mu_0$ = central surface brightness, $\rho_0$ = central density.\\
$^a$: Tabulated values for G1 represent the range quoted in the literature. 
See discussion in main text for details.\\
$^b$: data from the McMaster catalog (Harris 1996)\\
$^c$: Computed from $r_c$ and $r_t$.
}
\tablehead{   & Richstone & Brodie & G1$^a$ & $\omega$ Cen$^b$}
\startdata
$V_0$         & $19.190\pm0.005$     & $19.173\pm0.005$ & -    & -    \\
$(V-I)_0$     & $1.038\pm0.006$      & $1.050\pm0.006$  & -    & -    \\
$M_V$         & $-10.9$              & $-10.9$          & $-10.9$  & $-10.24$ \\
$[$Fe/H$]$    & \multicolumn{2}{c}{$-1.15\pm0.2$}& $-1.2$\ldots$-0.7$ & $-1.62$ \\
$c$           & \multicolumn{2}{c}{$200\pm100$}         & 166\ldots380   & 17.4  \\
minor/major   & $0.62\pm0.02$        & $0.64\pm0.01$    & $\sim0.75$ & 0.81 \\  
P.A.          & $-40.9\deg\pm1.1$    & $-38.3\deg\pm1.0$ &  -   &  -   \\
FWHM          & $0\farcs027\pm0\farcs006$ & $0\farcs029\pm0\farcs007$ & - & - \\
              & $1.27\pm0.29$ pc     & $1.39\pm0.32$ pc & - & -   \\
$r_c$ (pc)    &  $0.52\pm0.12$       & $0.57\pm0.13$ & 0.52\ldots0.63  & 3.8\\
$R_e$ (pc)$^c$ &  $3.47\pm0.33$      & $3.85\pm0.34$ & $4.5\pm0.5$ & 6.2  \\
$r_t$ (pc)    &  $95\pm27$           & $105\pm30$       & 90\ldots200  &  67  \\
$\mu_0$ ($V$ mag arcsec$^{-2}$) & $12.8\pm0.4$ & $13.0\pm0.4$ & 13.5 & 16.8 \\
$\rho_0$ (M$_{\odot}$ pc$^{-3}$) & $(4.7\pm2.8)\times10^5$ & $(3.5\pm2.0)\times10^5$ & $4.7\times10^5$ & $1.4\times10^3$ \\
\enddata
\end{deluxetable}

\end{document}